\documentclass[12pt ]{article}
\usepackage{latexsym} 
\usepackage{amsfonts}
\usepackage{amsmath}
\usepackage{amsthm}
\usepackage{amssymb}
\usepackage{graphicx}
\graphicspath{ {images/} }

\renewcommand{\theequation}
{\arabic{section}.\arabic{equation}}

\makeatletter
\def\eqnarray{ \stepcounter{equation} \let\@currentlabel=\theequation
 \global\@eqnswtrue
 \global\@eqcnt\z@
 \tabskip\@centering
 \let\\=\@eqncr
 $$\halign to \displaywidth\bgroup\@eqnsel\hskip\@centering
 $\displaystyle\tabskip\z@{##}$&\global\@eqcnt\@ne
 \hfil$\displaystyle{{}##{}}$\hfil
 &\global\@eqcnt\tw@$\displaystyle\tabskip\z@{##}$\hfil
 \tabskip\@centering&\llap{##}\tabskip\z@\cr}
\makeatother

\makeatletter
\def\@arrayacol{\edef\@preamble{\@preamble \hskip .5\arraycolsep}}
\def\array{\let\@acol\@arrayacol \let\@classz\@arrayclassz
\let\@classiv\@arrayclassiv \let\\\@arraycr\def\@halignto{}\@tabarray}
\makeatother

\makeatletter
\newcounter{subeqncnt}
\def\thesubeqncnt{\alph{subeqncnt}}
\def\subequations{\begingroup%
   \stepcounter{equation}\edef\@tempa{\theequation}%
   \let\c@equation\c@subeqncnt\c@subeqncnt\z@
   \edef\theequation{\@tempa\noexpand\thesubeqncnt}}

\makeatother

\newcommand{\be}{\begin{equation}}
\newcommand{\ee}{\end{equation}}

\newcommand{\beqa}{\begin{eqnarray}}
\newcommand{\eeqa}{\end{eqnarray}}
\newcommand{\nn}{\nonumber}

\renewcommand{\theequation}{\thesection.\arabic{equation}}
\newcounter{subequation}[equation]
\makeatletter

\expandafter\let\expandafter\reset@font\csname reset@font\endcsname

\def\subeqnarray{\arraycolsep1pt
    \def\@eqnnum\stepcounter##1{\stepcounter{subequation}%
        {\reset@font\rm(\theequation\alph{subequation})}}
\jot5mm     \eqnarray}

\makeatother

\def\CN {{\cal N}}

\def\half{\frac{1}{2}}

\def\alphadot{\dot {\alpha}}
\def\betadot{\dot {\beta}}
\def\e{{\varepsilon}}

\begin{document}

\setlength{\baselineskip}{7mm}
\begin{titlepage}
\begin{flushright}

{\tt NRCPS-HE-01-2017} \\

\end{flushright}

\begin{center}
{\Large ~\\{\it  Yangian and SUSY symmetry of High Spin 
Parton Splitting Amplitudes
in Generalised Yang-Mills Theory\\

\vspace{1cm}

}

}

\vspace{1cm}


{\sl Roland Kirschner$^{1}$  and  George Savvidy$^{2}$

\bigskip
\centerline{${}^1$ \sl Institut f\"ur Theoretische Physik, Universit\"at Leipzig} \centerline{\sl Augustusplatz 10, D-04109 Leipzig, Germany}
\bigskip
\centerline{${}^2$ \sl Institute of Nuclear and Particle Physics}
\centerline{ \sl Demokritos National Research Center, Ag. Paraskevi,  Athens, Greece}
\bigskip
}
\end{center}
\vspace{30pt}

\centerline{{\bf Abstract}}

We have calculated the high spin parton splitting amplitudes postulating the 
Yangian symmetry of the scattering amplitudes for
tensorgluons. The resulting splitting amplitudes  coincide with 
the earlier calculations, which were based on the BCFW recursion relations. 
The resulting formula unifies all known splitting probabilities found earlier 
in gauge field theories. It describes splitting probabilities for integer and half-integer spin particles. 
  We also checked that the splitting probabilities fulfil generalised 
Kounnas-Ross $\CN=1$ supersymmetry  relations  
hinting to the fact that the underlying theory can be formulated in an explicit supersymmetric  manner.

\vspace{12pt}

\noindent

\end{titlepage}



\pagestyle{plain}

\section{\it Introduction}

In the recent articles \cite{Savvidy:2014hha,Savvidy:2014uva,Savvidy:2015jgv} one of the authors (G.S.) considered a possibility that inside a proton and,
more generally, inside hadrons there could be
additional partons - tensorgluons, which could carry a part of the proton momentum and its spin.  The tensorgluons
have zero electric charge, like gluons, but have a larger spin \cite{Savvidy:2005fi,Savvidy:2005zm,Savvidy:2005ki,Savvidy:2010vb} and define  
asymptotically free fields similar to the standard Yang-Mills theory 
\cite{Gross:1973ju,Gross:1974cs,Politzer:1973fx}.

To describe the creation of tensorgluons and their density distribution  inside
the proton one should know the splitting amplitudes of gluons into tensorgluons. The corresponding amplitudes and the generalised 
DGLAP equations \cite{ Altarelli:1977zs,Dokshitzer:1977sg,
Gribov:1972ri,Gribov:1972rt, Lipatov:1974qm} which take into
account  the processes of emission of tensorgluons by gluons were derived in 
\cite{Savvidy:2014hha,Savvidy:2014uva,Savvidy:2015jgv}.

If the tensorgluons are created inside the proton one should also take 
into account the interaction of  tensorgluons 
of different spins between themselves.  These can be described in terms of
splitting probabilities $P^{h_C}_{h_B h_A}$. The full set of splitting
probabilities $P^{h_C}_{h_B h_A}$ - the kernels of the generalised 
DGLAP equations, 
describing the decay of tensorgluon of helicity $h_A$
 into two tensorgluons of helicities  $h_B$ and $h_C$ 
where derived in \cite{Savvidy:2014uva,Savvidy:2015jgv}. These splitting 
probabilities $P^{h_C}_{h_B h_A}$  fulfil very
general symmetry relations found earlier in \cite{ Altarelli:1977zs,Dokshitzer:1977sg, Gribov:1972ri,Gribov:1972rt, Lipatov:1974qm}.

Our aim in this article is to suggest alternative derivation 
of the splitting probabilities for tensorgluons postulating 
the infinite dimensional Yangian symmetry of the scattering amplitudes of the
tensorgluons \cite{Drummond:2008vq,ArkaniHamed:2010kv,ArkaniHamed:2012nw,Chicherin:2013sqa,Chicherin:2013ora,Fuksa:2016tpa}. The splitting probabilities calculated 
within the Yangian symmetry approach coincide with the earlier 
calculations based on the BCFW relations and hinting to the high symmetry of the generalised 
Yang-Mills theory amplitudes reminiscent to the symmetries  discovered   
in Yang-Mills theory  \cite{Drummond:2009fd,Drummond:2008vq}.  The splitting probabilities in this maximally symmetric 
representation have the following form:
\be\label{kernelsym0}
P^{h_C}_{h_B h_A}  = {1 \over z^{2 \eta  h_B -1} (1-z)^{2 \eta  h_{C} -1}    },
~~~h_C + h_B + h_A =\eta =\pm 1.
\ee
The formula describes all known splitting probabilities found earlier 
in QFT (\ref{setofquarkgluon}) and the generalised Yang-Mills theory 
(\ref{setoftensorgluonpolariz1}), (\ref{setoftensorgluonpolariz2}). 
This is a surprising 
and encouraging result because such a high symmetry was not 
explicitly implemented into the initial formulation. It was also 
interesting to check if the splitting probabilities (\ref{kernelsym0}) fulfil the 
generalised Kounnas-Ross supersymmetry  relations \cite{Kounnas:1982de,Jones:1983eh}.  As we shall demonstrate, the splitting probabilities 
 (\ref{kernelsym0}) fulfil the generalised $N=1$ SUSY relations (\ref{tworelation})
hinting to the fact that the underlying theory can be formulated in an explicit supersymmetric  manner \cite{Antoniadis:2011re}.

The present paper is organised as follows. In section two the basic formulae for
splitting probabilities and their symmetry relations are recalled, definitions  and notations are specified and generalised evolution equations for the tensorgluons
are presented. In section three we formulate the $s\ell_4$ Yangian symmetric 
amplitudes and extract the corresponding splitting amplitudes in the
collinear limit. In section four we derive the generalised $\CN=1$ Kounnas-Ross SUSY relations  and get convinced that they are fulfilled 
by the tensorgluons splitting probabilities (\ref{kernelsym0}). In section five we obtained the
$s\ell_2$ Yangian maximally symmetric representation of the tensorgluons splitting 
probabilities. In conclusion we summarise the results.

\section{\it Interaction Vertices and Splitting Probabilities }

In the generalised Yang-Mills theory \cite{Savvidy:2005fi,Savvidy:2005zm,Savvidy:2005ki,Savvidy:2010vb}
all interaction vertices
between high-spin particles have {\it dimensionless coupling constants},
which means that the helicities $h_i, i=1,2,3$ of the interacting particles in the vertex are
constrained  by the relation  
\be\label{consraint}
h_1+h_2+h_3= \pm 1~,
\ee
because the dimensionality
of the three-particle  vertex  $M_3(1^{h_1} ,2^{h_2},3^{h_3} )$  is $[mass]^{D=\pm(h_1+h_2+h_3)}$ \cite{Georgiou:2010mf,Benincasa:2007xk} and the 
condition (\ref{consraint}) means that the vertex has dimension of mass, as it 
is in the standard Yang-Mills theory \cite{Savvidy:2014hha,Savvidy:2014uva,Savvidy:2015jgv}. 
Therefore {\it on-mass-shell} interaction vertex between
massless tensorgluons   has the following form \cite{Savvidy:2014uva,Savvidy:2015jgv,Georgiou:2010mf,Benincasa:2007xk}:
\beqa\label{dimensionone1}
M_3(1^{h_1} ,2^{h_2},3^{h_3} ) &=&  g f^{abc} <1,2>^{-2h_1 -2h_2 -1} <2,3>^{2h_1 +1} <3,1>^{2h_2 +1},~~
h_3= -1 - h_1 -h_2, \nn\\
M_3(1^{h_1} ,2^{h_2},3^{h_3} ) &=&  g f^{abc} [1,2]^{2h_1 +2h_2 -1} [2,3]^{-2h_1 +1} [3,1]^{-2h_2 +1},~~~~~~h_3= 1 - h_1 -h_2,
\eeqa
where g is the YM coupling constant and $f^{abc}$ are the structure constants
of the internal gauge group G\footnote{
In subsequent equations we shall not write the factor $g f^{abc}$ explicitly. It is also understood that in a spinor representation of the on-mass-shell three-particle interaction vertices (\ref{dimensionone1})
the particle momenta are complexly deformed \cite{Benincasa:2007xk,Dixon:1996wi,Parke:1986gb,Witten:2003nn,
Britto:2005fq,Cachazo:2004kj,ArkaniHamed:2008yf,Berends:1988zn,Mangano:1987kp}. The alternative expressions for the three-particle interaction vertices can be found in \cite{Bengtsson:1983pd,Bengtsson:1983pg,Bengtsson:1986kh,Metsaev:2005ar,Adamo:2016ple}.}.
Considering the interaction vertex of the tensorgluons
of  helicities $h_A= \pm r$  and of helicities $h_C=  \pm s$, one can find from (\ref{consraint}) that the third particle helicity can take two values: 
$h_B =  \pm(s-r-1)   $, $s \geq 2r+1$ and
$h_B =  \pm(s-r+1) $, $s \geq 2r-1$,  while $r=1,2,3...$.

\begin{figure}
\includegraphics[width=10cm]{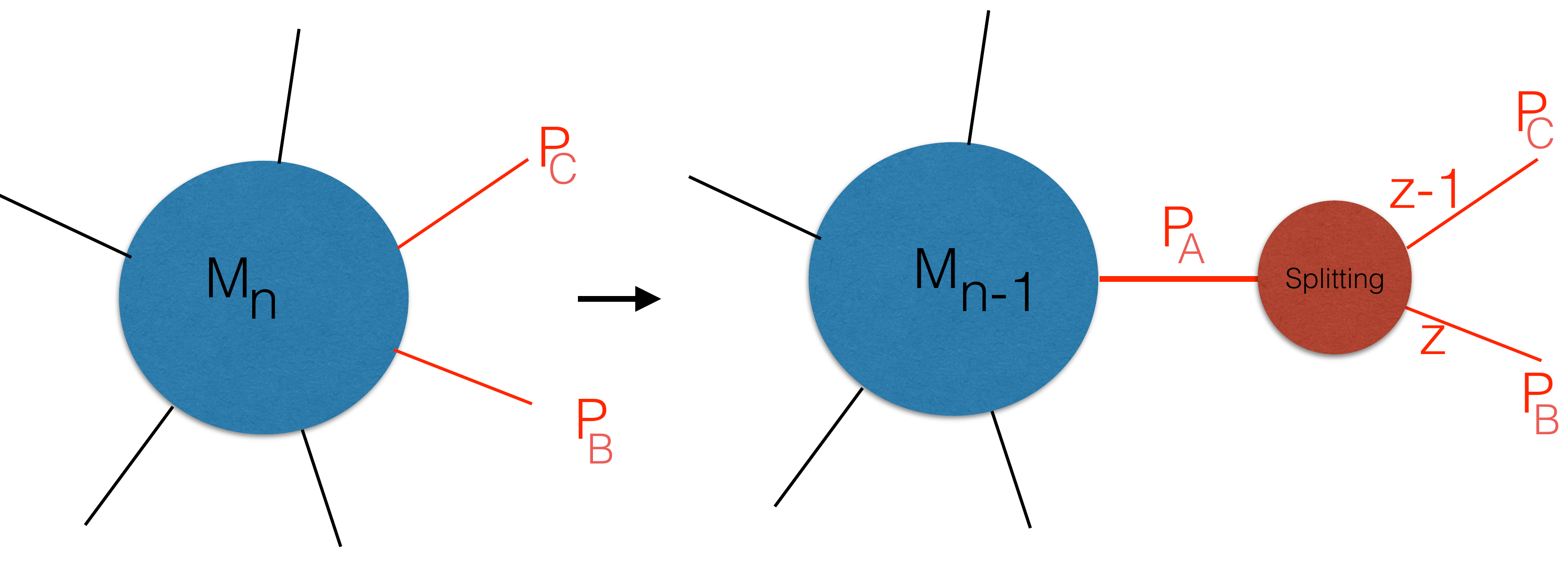}
\centering
\caption{The scattering amplitudes $M_n$ involving  tensorgluons
can be used to extract the splitting amplitudes 
$Split ( h_B,h_C, h_A)$ considering the limit when two neighbouring particles become collinear, $p_B  \parallel p_C$, ~
$p_B = z p_A,~p_C = (1-z) p_A$, ~ $p^2_A \rightarrow 0$ and $z$ describes the
longitudinal momentum sharing with the corresponding behaviour of spinors
$
\lambda_B = \sqrt{z} \lambda_A,~~~\lambda_C = \sqrt{1-z} \lambda_A .
$
  }
\label{fig1}
\end{figure}

\begin{figure}
\includegraphics[width=9cm]{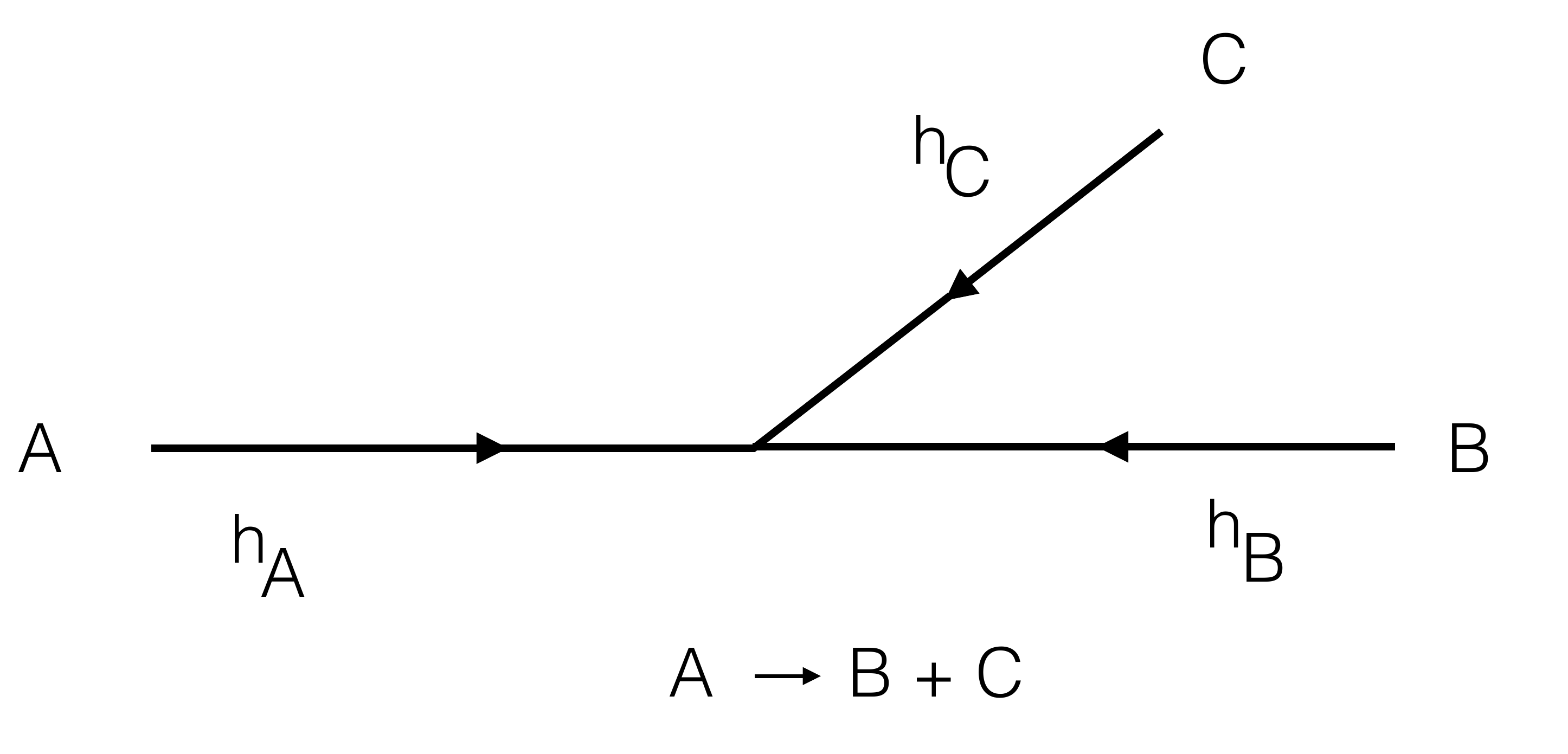}
\centering
\caption{The decay of the tensorgluon A into
tensorgluons B and C. The arrows
show the directions of the helicities.
The corresponding splitting probabilities are defined as $P^{h_C}_{h_B h_A}$.  }
\label{fig2}
\end{figure}

Using these vertices  one can compute the scattering amplitudes involving  tensorgluons
\cite{Savvidy:2014uva,Savvidy:2015jgv,Georgiou:2010mf,Antoniadis:2011rr} and extract splitting amplitudes 
$Split ( h_B, h_C,h_A)$ considering the limit when two neighbouring particles
become collinear, $p_B  \parallel p_C$, ~
$p_B = z p_A,~p_C = (1-z) p_A$, ~ $p^2_A \rightarrow 0$ and $z$ describes the
longitudinal momentum sharing with the corresponding behaviour of spinors
$
\lambda_B = \sqrt{z} \lambda_A,~~~\lambda_C = \sqrt{1-z} \lambda_A 
$
\cite{Dixon:1996wi,Parke:1986gb,Berends:1988zn,Mangano:1987kp,Antoniadis:2011rr} (see Fig. \ref{fig1}).   The residue of the collinear
pole in square  gives Altarelli-Parisi splitting probability $P(z)$
\cite{Dixon:1996wi,Parke:1986gb,Berends:1988zn,Mangano:1987kp,Antoniadis:2011rr}:
\be\label{AltarelliParisi}
P^{h_C}_{h_Bh_A}(z)= C_2(G)  ~ \vert Split ( h_B,h_C, h_A) \vert^2 ~ s_{BC},
\ee
where $s_{BC}=2 p_B \cdot p_C= <B,C>[B,C]$ (see Fig. \ref{fig2}).
The invariant operator $C_2$ for the representation R is defined by the equations
$ t^a t^a  = C_2(R)~ 1 $ and $tr(t^a t^b) = T(R) \delta^{ab}$.

The same splitting probabilities can be extracted directly by considering   {\it of-mass-shell}
decay of the  particle A.  It describes the probability of  finding a particle B inside a particle A
with fraction z of the longitudinal momentum of A  and radiation of the third particle C
with fraction $(1-z)$ of the longitudinal momentum of A \cite{Altarelli:1977zs}:
\be\label{transverce}
P^{C}_{BA}(z)= {1\over 2} z (1-z)  \sum_{helicities}
{\vert M_{A \rightarrow B+C}\vert^2 \over p^{2}_{\perp}},
\ee
where a sum is over the helicities  of B and C and the average over the helicity of A if one is interested in unpolarised splitting probabilities.
The important properties
of the splitting functions are the symmetries \cite{ Altarelli:1977zs,Dokshitzer:1977sg,
Gribov:1972ri,Gribov:1972rt, Lipatov:1974qm} over exchange of the
particles $B \leftrightarrow C$  with
complementary momenta fraction
\be\label{symmetryrelations1}
P^{C}_{BA}(z)= P^{B}_{CA}(1-z)
\ee
and a crossing relation
\be\label{symmetryrelations2}
P^{C}_{AB}(z)= (-1)^{2h_A +2h_B +1} z P^{C}_{BA}({1\over z}),
\ee
which emerges because two splitting processes are connected
by time reversal $A  \leftrightarrow B$.

The splitting probabilities (\ref{AltarelliParisi}) were calculated in \cite{Savvidy:2014uva} by using a complex deformation $w$ of the momenta 
in the triple vertex   (\ref{dimensionone1})   without breaking 
the mass shell conditions:
\be\label{momenta}
p_1=(\omega z, w, iw, k z),~p_2=(\omega (1-z), -w, -iw, k (1-z)),~p_3=(\omega , 0, 0, k ).
\ee
The corresponding polarization vectors are to be taken in the form:
\be
e^+_1 = {1\over \sqrt{2} } ( {z \over \omega}, 1, -i, - {z \over k}),~
e^+_2 = {1\over \sqrt{2} } (- {z \over \omega}, 1, -i, - {z \over k}),~
e^-_3 = {1\over \sqrt{2} } ( 0, 1, -i, 0),
\ee
fulfiling the following relations 
$
p^2_1=p^2_2=p^2_3=p_1 p_2=p_2 p_3=p_3 p_1=p_1 e^+_1=p_2 e^+_2=p_3 e^-_3=0.
$
The spinor representation of the momenta (\ref{momenta}) will take the following form:
\beqa
&\lambda_1=(\sqrt{(\omega+k)z},0),~~~&\tilde{\lambda}_{\dot{1}}=(\sqrt{(\omega+k)z},{2w\over \sqrt{(\omega+k)z}}),\nn\\
&\lambda_2=(\sqrt{(\omega+k)(1-z)},0),~~~&\tilde{\lambda}_{\dot{2}}=(\sqrt{(\omega+k)(1-z)},-{2w\over \sqrt{(\omega+k)(1-z)}}),\nn\\
&\lambda_3=(\sqrt{(\omega+k)},0),~~~&\tilde{\lambda}_{\dot{3}}=(\sqrt{(\omega+k)},0).
\eeqa
It follows  that the invariant products
$
<1,2>=<2,3>=<3,1>=0
$
vanish and that 
\be\label{products}
[1,2]=-2w {1\over \sqrt{z(1-z)}},~~~[2,3]=2w {1\over \sqrt{1-z}},~~~[3,1]=2w {1\over \sqrt{z}}.
\ee
Let us consider the interaction vertices (\ref{dimensionone1}) of tensorgluons
of the spins $A=r$, $C=s$  and  $B =  s-r+1 $,  where $s \geq 2r-1$, $r=1,2,3...$. Using the  scalar products (\ref{products}) for the vertices (\ref{dimensionone1})
one can get:
\beqa\label{rssvertex11}
M_3(1^{-s} ,2^{+r},3^{s-r+1} )&\propto&  { [2,3]^{2s+1} \over [1,2]^{2s-2r+1}  [3,1]^{2r-1}}=
-2 w {z^s \over (1-z)^r}\nn\\
M_3(1^{-s} ,2^{s-r+1},3^{+r} )&\propto&  {[2,3]^{2s+1} \over [1,2]^{2r-1}  [3,1]^{2s-2r+1}}=
-2 w {z^s \over (1-z)^{s -r+1} }\nn\\
M_3(1^{+r} ,2^{s-r+1},3^{-s} )&\propto&  {[1,2]^{2s+1} \over [2,3]^{2r-1}  [3,1]^{2s-2r+1}}=
-2 w {1 \over z^r (1-z)^{s -r+1} }\nn\\
M_3(1^{+r} ,2^{-s},3^{s-r+1} )&\propto&  {[3,1]^{2s+1} \over [2,3]^{2r-1}  [1,2]^{2s-2r+1}}=
-2 w {(1-z)^{s} \over z^r  }\nn\\
M_3(1^{s-r+1} ,2^{+r},3^{-s} )&\propto&  {[1,2]^{2s+1} \over [3,1]^{2r-1}  [2,3]^{2s-2r+1}}=
-2 w {1 \over z^{s-r+1} (1-z)^{r} }\nn\\
M_3(1^{s-r+1} ,2^{-s},3^{+r} )&\propto&  {[3,1]^{2s+1} \over [1,2]^{2r-1}  [2,3]^{2s-2r+1}}=
-2 w {(1-z)^{s} \over z^{s-r+1}  }.
\eeqa
These amplitudes can be written in a unified form as 
\be\label{rssvertex12}
M_3(h_B,h_C,h_A) \propto {-2 w \over z^{h_B} (1-z)^{h_C} }, ~~h_B+h_C +h_A=1.
\ee
Considering the transversal momentum $p_{\perp}$  in (\ref{transverce})
to be proportional to the  deformation parameter $p_{\perp} \propto w$   
one can get the following expression  for splitting  probabilities:
\beqa
P(z)= {1\over 2}z(1-z) \vert M_3 \vert^2 {1\over \vert w \vert^2} \nn,
\eeqa
and then, by using (\ref{rssvertex11}), the following set of splitting probabilities 
\cite{Savvidy:2014uva,Savvidy:2015jgv}:
\beqa\label{setoftensorgluonpolariz1} 
P^{s}_{s-r+1,r} &=& C_2(G) {(1-z)^{2s+1} \over z^{2s-2r+1}  },~~~~
P^{s-r+1}_{s,r} = C_2(G)  {z^{2s+1} \over (1-z)^{2s -2r+1} }\nn\\
P^{s}_{r,s-r+1} &=& C_2(G) {(1-z)^{2s+1} \over z^{2r-1}  },~~~~
P^{s-r+1}_{r,s} =  C_2(G){1 \over z^{2r-1} (1-z)^{2s -2r+1} }\nn\\	
P^{r}_{s,s-r+1}&=& C_2(G)  { z^{2s+1} \over (1-z)^{2r-1}   },~~~~
P^{r}_{s-r+1,s} = C_2(G) {1 \over z^{2s-2r+1} (1-z)^{2r-1} },
\eeqa
where $s \geq 2r-1$, $r=1,2,3...$.
The splitting probabilities for $A=r$, $C=s$  and  $B =  s-r-1 $ are: 
\beqa\label{setoftensorgluonpolariz2}
P^{s}_{s-r-1,r} &=& C_2(G) {  z^{2s-2r-1}  \over  (1-z)^{2s-1}},~~~~
P^{s-r-1}_{s,r} = C_2(G)  {(1-z)^{2s -2r-1} \over  z^{2s-1} }\nn\\
P^{s}_{r,s-r-1} &=& C_2(G) {z^{2r+1}  \over (1-z)^{2s-1}  },~~~~
P^{s-r-1}_{r,s} =  C_2(G) z^{2r+1} (1-z)^{2s -2r-1} \nn\\	
P^{r}_{s,s-r-1}&=& C_2(G)  { (1-z)^{2r+1}   \over  z^{2s-1} },~~~~
P^{r}_{s-r-1,s} = C_2(G)   z^{2s-2r-1} (1-z)^{2r+1} ,
\eeqa
where $s \geq 2r+1$  $r=1,2,3...$ .    The  expressions
(\ref{setoftensorgluonpolariz1}), (\ref{setoftensorgluonpolariz2})
describe all possible splitting probabilities
corresponding to the interaction vertices  
of the generalised YM theory  \cite{Savvidy:2014hha,Savvidy:2014uva,Savvidy:2015jgv,Savvidy:2005fi,Savvidy:2005zm,Savvidy:2005ki}
and can be written in a unified form as 
\be\label{rssvertex13}
P^{h_C}_{h_B h_A} = {C_2(G) \over z^{2h_B-1} (1-z)^{2h_C-1} }, ~~h_B+h_C +h_A=1.
\ee
The splitting probabilities (\ref{setoftensorgluonpolariz1}), (\ref{setoftensorgluonpolariz2}), (\ref{rssvertex13}) fulfil the symmetry realtions
(\ref{symmetryrelations1}), (\ref{symmetryrelations2}).
For completeness we shall present also quark and gluon
splitting probabilities \cite{Altarelli:1977zs}:
\beqa\label{setofquarkgluon}
P^G_{qq}(z) &=& C_2(R){1+z^2 \over 1-z },\nn\\
P^q_{Gq}(z) &=& C_2(R)[{1\over z} +{(1-z)^2 \over z }],\\
P^q_{qG}(z) &=& T(R)[z^2 +(1-z)^2], \nn\\
P^G_{GG}(z) &=&  C_2(G)\left[{1 \over z(1-z)}+ {z^4 \over z(1-z)}+{(1-z)^4 \over z(1-z)}\right],\nn
\eeqa
where $C_2(G)= N, C_2(R)={N^2-1  \over  2 N},  T(R) = {1  \over  2}$ for the SU(N) groups.
 
Using the splitting probabilities
for the tensorgluons (\ref{setoftensorgluonpolariz1}), (\ref{setoftensorgluonpolariz2})
one can derive the evolution equations which will take into account
a possible emission of tensorgluons  in a proton \cite{Savvidy:2014uva,Savvidy:2015jgv}. Introducing
the corresponding densities $T_s(x, t)$ of tensorgluons (summed over colours)
inside a proton in the $P_{\infty}$ frame  one can derive
the integro-differential equations that describe the $Q^2$ dependence
of parton densities in this general case. They are \cite{Savvidy:2014uva,Savvidy:2015jgv}:
\beqa\label{evolutionequation}
{d q_i(x,t)\over dt} &=& {\alpha(t) \over 2 \pi} \int^{1}_{x} {dy \over y}[\sum^{2 n_f}_{j=1} q_j(y,t)~
P_{q_i q_j}({x \over y})+ G(y,t)~ P_{q_i G}({x \over y})] ,\\
{d G(x,t)\over dt} &=& {\alpha(t) \over 2 \pi} \int^{1}_{x}
{dy \over y}[\sum^{2 n_f}_{j=1} q_j(y,t)~
P_{G q_j}({x \over y})+ G(y,t) ~P_{G G}({x \over y})+ \sum_{s} T_s(y,t) ~P_{G T_s}({x \over y}) ],\nn\\
{d T_r(x,t)\over dt} &=& {\alpha(t) \over 2 \pi} \int^{1}_{x} {dy \over y}[
G(y,t)~ P_{T_r G}({x \over y}) +  \sum_{s} T_{s}(y,t)~ P_{T_{r} T_{s} }({x \over y})].\nn
\eeqa
The $\alpha(t)$ is the running coupling constant ($\alpha = g^2/4\pi$).
In the leading logarithmic approximation $\alpha(t)$ is of the form
\be\label{strongcouplingcons}
{\alpha \over \alpha(t)} = 1 +b ~\alpha ~t~~,
\ee
where $\alpha = \alpha(0)$ and $b$ is the one-loop Callan-Symanzik coefficient.
The densities of the quarks and of gluons are changing
because of  the standard radiation processes, the density of
tensorgluons changes because there are transitions 
between them through the
splittings which are described by the probabilities (\ref{setoftensorgluonpolariz1}), (\ref{setoftensorgluonpolariz2}).
In the next section we shall derive the splitting amplitudes for 
the tensorgluons postulation the Yangian symmetry of the amplitudes.

\section{\it Yangian Symmetry of Parton Splitting Amplitudes }
 
In this section we shall  present an alternative derivation 
of the splitting probabilities for tensorgluons postulating 
the $s\ell_4$ Yangian symmetry of the scattering amplitudes of the
tensorgluons \cite{Drummond:2009fd, Drummond:2008vq,Chicherin:2013sqa,ArkaniHamed:2010kv,ArkaniHamed:2012nw,Chicherin:2013ora,Ferro:2014gca,Fuksa:2016tpa}. As we shall demonstrate, the splitting amplitudes calculated 
within the Yangian symmetry approach coincide with 
(\ref{setoftensorgluonpolariz1}), (\ref{setoftensorgluonpolariz2}) 
 and hint to the high symmetry of the generalised 
Yang-Mills theory amplitudes reminiscent to the symmetries  discovered   
in Yang-Mills theory \cite{Drummond:2009fd, Drummond:2008vq,ArkaniHamed:2010kv,ArkaniHamed:2012nw,Chicherin:2013sqa,Chicherin:2013ora,Ferro:2014gca,Fuksa:2016tpa}. This is a surprising 
and encouraging result because such a high symmetry was not 
explicitly implemented into the initial formulation \cite{Savvidy:2014hha,Savvidy:2014uva,Savvidy:2015jgv}.

We shall derive the splitting amplitudes $Split(h_B,h_C,h_A)$ from the collinear limit of $s\ell_4$ Yangian
symmetric amplitudes \cite{Chicherin:2013sqa,Chicherin:2013ora,Fuksa:2016tpa}. The latter are defined as eigenfunctions of the monodromy operator of a $s\ell_4$ symmetric integrable spin chain, periodic with $N$
sites, composed of the appropriate $4\times 4$ $L$ matrix operators:
\be \label{YSC}
 T({\bf u}) = \prod_1^N L_i(u_i^+, u_i), \ \  T({\bf u}) M(1,...,N) =
E({\bf u}) M(1,...,N). \ee
The matrix elements of $L_i$ are operators being generators of the $s\ell_4$
algebra and acting on the variables in $M(1,...,N)$
associated with the point $i=1, ..., N$. We use the helicity representation,
where these variables are the Weyl spinor components $ \lambda_{i,\alpha}, \bar
\lambda_{i, \alphadot}, \alpha=1,2, \alphadot=1,2 $. The dependence on the
variables at $i$ is homogeneous in the sense that the dilatation of the
spinors $\bar \lambda_i \to t \bar \lambda_i, \ \lambda_i \to t^{-1}
\lambda_i $ implies for the correlation $M(1, ...,N) \to t^{2h_i-2} M(1,
...,N)$. The degree of homogeneity is related to the spectral parameters as
$u_i^+  =  u_i + 2h_i-2 $. In the helicity representation the 
matrix elements of $L(u^+, u) \to  Iu + L(0)$ are
$$ L(0) =
\begin{pmatrix}  
L_{\alpha,\beta} & L_{\alpha, \betadot} \cr
L_{\alphadot, \beta} & L_{\alphadot, \betadot}
\end{pmatrix},
$$
$$ L_{\alpha,\beta} = - \lambda_{\alpha} \partial_{\beta}, ~ L_{\alpha, \betadot}
= - \lambda_{\alpha} \bar \lambda_{\betadot} , ~
L_{\alphadot, \beta} = \partial_{\beta} \bar \partial_{\alphadot}, ~
L_{\alphadot, \betadot} = \bar \partial_{\alphadot} \bar \lambda_{\betadot} . $$
The N-point Yangian symmetric correlator can be identified with the scattering
amplitudes, where a particle state related to the leg $i$ is represented
by the spinors in the known way, in particular, its momentum is $
p_{i, \alpha,\alphadot} = \sigma^{\mu}_{\alpha,\alphadot} p_{i, \mu} =
\bar \lambda_{i, \alphadot} \lambda_{i, \alpha} $ and the particle type is
fixed by substituting the physical helicity value for the parameter $h_i$. 

We shall consider the particular solutions of the $s\ell_4$ invariant amplitude with $N=5$ and $N=4$ particles,
\beqa \label{M52}
&M_{5} =  \delta^{(4)} ( \sum_1^5 \lambda_{p,\alpha} \bar \lambda_{p,
\alphadot} ) \nn\\
&<12>^{-1+2h_3+2h_5} <2 3>^{-1 + 2h_1 + 2h_4} <34>^{-1+2h_2+2h_5} 
<45>^{-1 +2h_1+2h_3} <51>^{-1+ 2h_2+2h_4}=\nn\\
&= \delta^{(4)} ( \sum_1^5 \lambda_{k,\alpha} \bar \lambda_{k,
\alphadot} ) \ \ \prod_1^5 <i-1, i>^{-1 +2h_{i-2}+2h_{i+1}} ~,
\eeqa
where the helicities obey the constraint 
$ \sum_1^5 h_i = 1 $ and $h_{i+5}=h_i$.  In the case $N=4$ we obtain
\be \label{M42}
M_{4} = \delta^{(4)} ( \sum_1^4 \lambda_{k,\alpha} \bar \lambda_{k,
\alphadot} ) 
 \left( \frac{<12> <34>}{<23><41>} \right)^{\e} 
<12>^{-1} <23>^{1-2h_2} <34>^{-1 +2h_1+2h_2} <41>^{1-2h_1} ,
\ee 
where only two helicities are independent
$ h_3 = -h_1, \ \ h_4 = - h_2 $ and $\e $ remains as a free parameter. 
The expressions (\ref{M52})(\ref{M42}) are related to the one formulated in 
\cite{Ferro:2014gca,Bargheer:2014mxa} for the deformed Grassmannian of $\mathcal{N} =4$ 
super Yang-Mills scattering amplitudes.

In order to extract the splitting amplitudes
for tensorgluons we shall consider the collinear limit of $M_{5}$ in  
equation \eqref{M52}
(see Fig.\ref{fig1}):
$$
 p_i \to z p, \ \ p_{i+1} \to (1-z) p, \ \ \lambda_i \to \sqrt z
\lambda_p, \ \ \lambda_{i+1} \to \sqrt{1-z} \lambda_p ~. $$
For the products of helicity variables this means that
$$ <i, i+1> \to 0 , \ \ <i-1, i> \to <i-1, p> \sqrt z, \ \ 
 <i+1, i+2> \to <p, i+2> \sqrt{1- z}.
$$
The factorisation with a one-particle intermediate state
occurs if the exponent at $<i, i+1>$ in (\ref{M52}) is $-1$, that is 
$h_{i-1} + h_{i+2} = 0$ and from the constraint  $ \sum_1^5 h_i = 1$  
it follows then that
\be \label{1pi}
h_{i-2} +  h_i + h_{i+1}  = 1.
\ee
In the collinear limit we have (omitting the energy-momentum delta
distribution)
$$ 
M_{5} ~\rightarrow ~ (<i-2,i-1>)^{-1+2h_i+2h_{i+2} }  (<i-1,p>\sqrt
z)^{-1+2h_{i-2}+2h_{i+1}}  <i,i+1>^{-1} 
$$ 
$$ 
(<p,i+2>\sqrt{1-z})^{-1 + 2h_i+2h_{i-2}} 
<i+2, i-2>^{-1 +2h_{i+1} +2h_{i-1}}. 
$$
The last expression has a factorised form:
\be \label{fact}
 M_{5} \to  M_{4} <i,i+1>^{-1} Split(h_i,h_{i+1},h_p), 
\ee
where the first factor coincides with the 4-point amplitude \eqref{M42}:
$$  M_{4} = <i-2,i-1>^{-1+2h_i+2h_{i+2} }
<i-1,p>^{-1+2h_{i-2}+2h_{i+1}} $$ $$
<p,i+2>^{-1 + 2h_i+2h_{i-2}}
<i+2, i-2>^{-1 + 2h_{i+1} +2h_{i-1}}
= $$ 
$$
\left ( \frac{<i-2,i-1> <p, i+2>}{<i-1,p> <i+2,i-2>} \right)^{2h_i-2}
<i-2,i-1>^{1-2h_{i-1}} <i-1,p>^{-1} $$ 
$$ 
<p, i+2>^{1+ 2 h_{i-2} } <i+2, i-2>^{-1-2h_{i-2} +2h_{i-1}},
$$
if one relables the indices   $1,2,3,4 \to i-1, p, i+2, i-2$ and takes 
$\e \to 2 -2h_i $.  The last factor is 
\beqa
 (\sqrt z)^{-1+2h_{i-2}+2h_{i+1}} (\sqrt{1-z})^{-1 + 2h_i+2h_{i-2}} =
\sqrt{ z (1-z)}  z^{- h_{i}} ( 1-z)^{- h_{i+1}}, 
\eeqa
where we  used the  relations $2 h_{i+1} +2 h_{i-2}=2 - 2 h_{i}$ and 
$2h_{i}+2h_{i-2}=2 -2 h_{i+1}$, which follow from (\ref{1pi}). 
Thus we were able to extract the splitting amplitude for tensorgluons 
which has therefore the following elegant form: 
\be 
Split(h_i,h_{i+1},h_p) = \frac{\sqrt{z(1-z)} }{z^{h_i} (1-z)^{h_{i+1}} } .
\ee
The helicity of the intermediate state is denoted by $h_p = h_{i-2}$ and
obeys the relation \eqref{1pi} 
$
 h_p + h_i + h_{i+1} = 1.
$
This condition coincides with the dimensionless condition on the interaction vertices 
of the generalised Yang-Mills theory (\ref{consraint}), and here it appears as a consequence of the conformal invariance of the three-particle interaction 
vertices.  If one starts instead with the amplitudes corresponding to the
parity reflected particles, then we shall obtain that the splitting amplitudes 
fulfil the alternative constrain $\sum h_i = -1$. Introducing the sign symbol 
$\eta= \pm 1$ we can formulate both cases in one expression as 
\be \label{split} 
Split(h_i,h_{i+1},h_p) = \frac{\sqrt{z(1-z)} }{z^{\eta h_i} (1-z)^{\eta
h_{i+1}}}, \ \ h_i+h_{i+1}+h_p = \eta,  
  \ee
and for the splitting probability (\ref{AltarelliParisi}) we shall get
\be\label{kernelsym}
P^{h_C}_{h_B h_A}  = {1 \over z^{2 \eta  h_B -1} (1-z)^{2 \eta  h_{C} -1}    },
~~~h_C + h_B + h_A =\eta.
\ee
It is interesting to notice that the splitting probabilities (\ref{setoftensorgluonpolariz1}),(\ref{setoftensorgluonpolariz2}) and
(\ref{kernelsym}) can be represented in the following 
symmetric form: 
\be \label{phi4}
P^{h_C}_{h_B h_A} = {k_B k_C k_A \over k_B^{2 \eta h_B}
k_C^{2 \eta h_C} k_A^{ 2\eta h_A}},
\ee
where the one-dimensional light-cone momenta are  defined as in (\ref{momenta}):
$
k_A=1,~k_B=z,~k_C=(1-z)
$. In the subsequent sections this expression will be rigorously derived as the light-cone momentum 
factor of the
$s\ell_2$ Yangian symmetric amplitude in the two-dimensional 
helicity representation $h_i$ of the solution of the 
$s\ell_2$ version of the equation (\ref{YSC})  \cite{Fuksa:2016tpa}.

\section{\it SUSY Symmetry of the Splitting Amplitudes}

In the supersymmetric QCD the splitting amplitudes and probabilities fulfil supersymmetric relations 
which were established  in \cite{Kounnas:1982de,Jones:1983eh,Bukhvostov:1985rn}. These $\CN=1$ Kounnas-Ross relations are between  splitting probabilities $P_{BA}$ of the  members of the supersymmetric multiplets consisting of the matter supermultiplet of quarks $(q_i)$ and squarks $(s_i,t_i)$ and of the vector 
supermultiplet  of gluons $(G)$ and gluinos $(\lambda)$:
\beqa
P_{GG} + P_{\lambda G}=P_{G \lambda} + P_{ \lambda \lambda}\nn\\
P_{Gq} + P_{\lambda q}=P_{G s} + P_{  \lambda s}\nn\\
P_{qG} + P_{s G}=P_{q\lambda} + P_{ s \lambda}\nn\\
P_{qq} + P_{sq}=P_{qs} + P_{ss}.
\eeqa
The first relation is well known from the standard QCD   when the quarks are in the adjoint representations of SU(3) \cite{Dokshitzer:1977sg}.

It is interesting to check if the high spin evolution kernels  $P^{h_C}_{h_B h_A}$ fulfil generalised $\CN=1$ supersymmetry  relations.  As we shall demonstrate, the splitting probabilities  fulfil the $N=1$ SUSY relations 
hinting to the fact that the underlying theory can be formulated in an explicit supersymmetric  manner. Indeed, considering the supermultiplets 
$(1,1/2)$ and  $(s,s-1/2)$ we shall get the relations including the  gluons,   gluinos and  tensorgluons with their partners tensorgluionos:
\be\label{onerelation}
P^{1/2}_{s (s-1/2)} + P^{1}_{(s-1/2) (s-1/2)}= P^{1/2}_{(s-1/2)s} + P^{1}_{s s}.
\ee
For the corresponding polarisation  kernels  we shall have the following expressions (\ref{setoftensorgluonpolariz1}),(\ref{setoftensorgluonpolariz2}), (\ref{kernelsym}):
\beqa
P^{1/2}_{s^- (s-1/2)^+} =0,~~P^{1}_{(s-1/2)^-(s-1/2)^+} ={z^{2s}\over 1-z},~~P^{1/2}_{(s-1/2)^-s^+} =z^{2s},~~P^{1}_{s^-s^+} ={z^{2s+1}\over 1-z}; ~~~~~ ~~~~~~\\
P^{1/2}_{s^+ (s-1/2)^-} ={1\over z^{2s-1}},P^{1}_{(s-1/2)^+(s-1/2)^-} ={1\over z^{2s-2}(1-z)},P^{1/2}_{(s-1/2)^+s^-} =0,P^{1}_{s^+s^-} ={1\over z^{2s-1}(1-z)},\nn
\eeqa
and, as one can see, each set of these polarisation kernels  fulfils the $\CN=1$
relation (\ref{onerelation}). Let us also consider two arbitrary supermultiplets 
$(s,s-1/2)$ and $(r,r-1/2)$.  For these supermultiplets the $\CN=1$  SUSY
relation has the following generalised Kounnas-Ross form: 
\be\label{tworelation}
P^{s-r +1/2}_{r (s-1/2)} + P^{s-r+1}_{(r-1/2) (s-1/2)}= P^{s-r+1/2}_{(r-1/2)s} + P^{s-r+1}_{r s}.
\ee
Calculating the corresponding splitting kernels we shall get
\beqa
&P^{(r-s+1/2)^+}_{r^- (s-1/2)^+} =0,~~&P^{(r-s+1)^+}_{(r-1/2)^-(s-1/2)^+} ={z^{2r}\over (1-z)^{2r-2s+1}},~\nn\\
&P^{(r-s+1/2)^+}_{(r-1/2)^-s^+} ={z^{2r}\over (1-z)^{2r-2s}},~~&P^{(r-s+1)^+}_{r^-s^+} ={z^{2r+1}\over (1-z)^{2r-2s+1}};\nn\\
&P^{(s-r+1/2)^+}_{r^+ (s-1/2)^-} ={1\over z^{2r-1}(1-z)^{2s-2r}},~~&P^{(s-r+1)^+}_{(r-1/2)^+(s-1/2)^-} ={1\over z^{2r-2}(1-z)^{2s-2r+1}},~\nn\\
&P^{(s-r+1/2)^+}_{(r-1/2)^+s^-} =0,~~&P^{(s-r+1)^+}_{r^+s^-} ={1\over z^{2r-1}(1-z)^{2s-2r+1}}
\eeqa
and, as one can see, both sets of polarisation kernels fulfil the supersymmetry 
relation (\ref{tworelation}). 

In the next section we shall consider the amplitudes which are 
the solution of the $s\ell_2$ version of Yangian symmetry (\ref{YSC}) obtained 
in  \cite{Fuksa:2016tpa}.  These amplitudes represent a longitudinal, two-dimensional reduction  \cite{Bjorken:1969ja} of the four-dimensional $s\ell_4$  symmetric amplitudes
and, as we shall see, the alternative derivation of the splitting probabilities will coincide with the one presented above but has the advantage to represent the 
results in a more symmetric form.

\section{\it $s\ell_2$ Symmetries of the splitting amplitudes}

We notice that the splitting amplitude can be regarded as a result of a
particular substitution in the function of 3 one-dimensional light-cone 
momenta $k_1,k_2,k_3, k_1+k_2+k_3=0$,
\be \label{phi3}
\phi(a_1,a_2,a_3;k_1,k_2,k_3) = (k_1 k_2 k_3)^{\half} k_1^{- \eta a_1}
k_2^{- \eta a_2} k_3^{- \eta a_3}, \ \ \ \sum a_i = \half \eta 
\ee
$$ 
Split(h_1, h_2,h_3;z) = \phi( h_1,h_2, h_3- \half \eta ; z,  1-z, -1). 
$$
The parton splitting probabilities are calculated as squares of the
corresponding splitting amplitudes. The helicities refer to ingoing momenta,
i.e. $h_1,h_2$ are opposite to their physical values in the decay
$3 \to 1+2$:
$$ P _{ h_1 \ h_3}^{h_2} (z) = 
Split^2(h_1,h_2,h_3;z) = \phi^2(h_1, h_2,
h_3 -\half \eta;-z,-1+z, 1). 
$$
The expressions for the parton splitting probabilities given in sect. 2
are reproduced. 

The simple expression for $\phi(a_1,a_2,a_3;k_1,k_2,k_3)$ results in
a number of trivial relations which result through
the above substitutions in  well known relations of the parton kernels
with obvious physical interpretations.
This expression can be obtained as the light-cone momentum factor in the
$s\ell_2$ Yangian symmetric 3-point function in the 2-dimensional analogon
of the helicity representation. The latter can be 
derived as a solution of the $s\ell_2$ version of (\ref{YSC}), as
explained in \cite{Fuksa:2016tpa}. The explicit form of the $L$ matrix
(in the case $\eta= +1$) is  $L(u^+,u) \to Iu + L(0)$
$$ L(0) = 
\begin{pmatrix}
 S^0 & S^- \cr
S^+ & -S^0  
\end{pmatrix},
$$ $$
S^0 = - k\partial_k, S^- = -k, S^+ = \frac{1}{k} ( k\partial_k +a-\half)
(k\partial_k-a-\half) $$ 
$S^b, b= 0, \pm$ obey the $s\ell_2$ Lie algebra relations.
Indeed, the 3-point function $\phi$ is determined by the following 
conditions:
$$ (S_1^b + S_2^b + S_3^b) \phi ~\delta(\sum k_i)= 0, \  b= 0, \pm. $$ 
We consider some relations for the symmetric 3-point function and 
their implications for the splitting probabilities.
The relation of parity symmetry for flipping all helicities is
obvious in this form, because 
$\phi(a_1,a_2,a_3;k_1,k_2,k_3)= \phi(-a_1,-a_2,-a_3;k_1,k_2,k_3)$
implies
$$ P _{ h_1 \ h_3}^{h_2} (z) = P _{ - h_1 \ - h_3}^{-h_2} (z).
$$
Further the crossing relations for the exchange of the
helicity labels at  $P _{ h_2 \ h_3}^{h_1} (z)$ follow easily from
$
\phi(a_1,a_2,a_3;k_1,k_2,k_3)=\phi(a_2,a_1,a_3;k_2,k_1,k_3),~~
 \phi(a_2,a_1,a_3;k_1,k_2,k_3)=\phi(a_1,a_2,a_3;k_2,k_1,k_3) $.
Indeed, the first relation results in 
$$ 
P _{ h_2 \ h_3}^{h_1} (z)= P _{ h_1 \ h_3}^{h_2} (1-z). 
$$
The second relation results in
$$  
P _{ h_3 \ h_1}^{h_2} (z) = \pm \ z  P _{ h_1 \ h_3}^{h_2}
(\frac{1}{z}) .
$$
The last relation is obtained  by substituting $ z= -\frac{k_1}{k_2}, 1-z=
- \frac{k_3}{k_1}$ and using $\eta(h_1+h_2+h_3) = \half,$ . 
As an intermediate step we rewrite 
 $\phi$ by using the constraints on the sum of momenta and the sum of
parameters $a_i$
as
$$ 
\phi(a_1,a_2,a_3;k_1,k_2,k_3) = (k_1k_2 k_3)^{\half} k_1^{-\half}
(\frac{k_2}{k_1})^{-\eta a_2} (-1-\frac{k_2}{k_1})^{-\eta a_3} .
$$ 
In this way we reproduce the well known crossing relations for the parton splitting
probabilities \cite{Altarelli:1977zs,Dokshitzer:1977sg}. 
In this representation we have supersymmetry relations due to momentum conservation.  
The shift of the parameter $a_i$ by $-\half \eta$ results in an extra factor
$k_i$,  therefore
\beqa
&& \phi^2(a_1-\half\eta, a_2,a_3; k_1,k_2,k_3) 
+ \phi^2(a_1,a_2-\half\eta,a_3; k_1,k_2,k_3)+  \nn\\
&&\phi^2(a_1, a_2,a_3-\half\eta; k_1,k_2,k_3) =
\phi^2(a_1,a_2,a_3;k_1,k_2,k_3) \ (k_1+k_2+k_3) = 0 .
\eeqa
We rewrite this equation in terms of the splitting amplitudes as
$$ 
Split^2(h_1-\half\eta,h_2, h_3+\half\eta;z) +
Split^2(h_1,h_2-\half\eta, h_3+\half\eta;z) + Split^2(h_1+h_2,h_3;z)= 0 
$$
and obtain a non-trivial relation for the splitting probabilities
which can be related to supersymmetry, because it involves parton helicities
differing by $\half$:
 \be \label{susya}
P_{h_1 h_3}^{h_2}(z) -
P_{ h_1 -\half \eta , h_3+\half \eta}^{h_2}(z) -
P_{h_1, h_3+\half\eta}^{h_2- \half\eta}(z) = 0. 
\ee
The signs appear in turning from the incoming convention for the momenta to
the physical situation $3 \to 1 + 2$.
By the substitution $h_3 +\eta \half \to  h_3,~ h_1 - \eta \half \to h_1$ 
we obtain another form of the same relation:
\be \label{susyb} 
P_{h_1, h_3}^{h_2}(z) +
P_{h_1+ \half\eta, h_3}^{h_2 - \half \eta}(z) -
P_{h_1+ \half \eta, h_3-\half\eta}^{ h_2}(z) = 0. 
\ee
The parton scale evolution involving the doublets of helicities
$(h_3, h_3-\half)$, $(h_1, h_1 +\half)$ 
is supersymmetric if the following relation holds:
\be \label{susy}
 P_{h_1, h_3-\half} + P_{h_1, h_3} = P_{h_1+\half, h_3-\half} +
P_{h_1+ \half, h_3}.
\ee
Here the helicities $h_2$ of the exchange parton are summed over  
$$ P_{h_1, h_3} = P_{h_1, h_3}^{ +1 - h_1-h_3} + P_{h_1,h_3}^{-1- h_1-h_3} 
= $$ $$
\frac{z}{1-z} \left [ \frac{z^{2h_1} }{  (1-z)^{2(h_3+h_1)} } +  \frac{
(1-z)^{2(h_1+h_3)}}{z^{2h_1}} \right ] .
$$
In the above supersymmetry relation thus the helicity values 
for $h_2$ are
$  h_+ = +1-h_1-h_3, h_- =-1-h_1-h_3, h_+ \pm \half,  h_- \pm \half $.
Substituting this into the Susy relation (\ref{susy}) we would have
$$ (P_{h_1,h_3- \half}^{ h_- +\half} + P_{h_1,h_3- \half}^{h_+
+\half}) + (P_{h_1 , h_3}^{h_-} + P_{h_1,h_3}^{ h_+}) = $$ $$
(P_{h_1+ \half, h_3- \half}^{ h_- } + P_{h_1+\half , h_3- \half}^{h_+})
+
(P_{h_1 +\half,h_3}^{ h_- -\half} + P_{h_1+\half , h_3}^{h_+- \half}).  
$$
We show that this cannot be valid without restriction.
We write (\ref{susya}) for $\eta=-1$ and (\ref{susyb})
for $\eta=+1$. 
$$ P_{h_1, h_3}^{h_2}(z) -
P_{h_1 +\half ,h_3-\half }^{h_2}(z) -
P_{h_1, h_3-\half}^{h_2+ \half\eta}(z) = 0 , \ \ \  
h_2 = -1 -h_1-h_3 = h-$$
$$P_{h_1, h_3}^{h_2}(z) +
P_{h_1+ \half, h_3}^{h_2 - \half }(z) -
P_{h_1+ \half , h_3-\half}^{ h_2}(z) = 0, \ \ \ 
h_2 = +1 -h_1-h_3 = h_+$$
The sum of these relations reproduces the supersymmetry
relation (\ref{susy}) if the contributions with $h_2 = h_- -\half $ and $h_2 = h_+
+\half $ are excluded.

 \section{\it Conclusion}

The aim of this article was to suggest an alternative derivation 
of the splitting probabilities for tensorgluons postulating 
the infinite dimensional Yangian symmetry of the scattering amplitudes of the
tensorgluons.  As we demonstrated, the splitting probabilities calculated 
within the Yangian symmetry approach coincide with the earlier 
calculations, which were based on the BCFW recursion relations and were hinting to the high symmetry of the generalised 
Yang-Mills theory amplitudes reminiscent to the symmetries  discovered   
in Yang-Mills theory.  
The splitting probabilities have the following highly symmetric and universal form:
\be\label{kernelsym1}
P^{h_C}_{h_B h_A}  = {1 \over z^{2 \eta  h_B -1} (1-z)^{2 \eta  h_{C} -1}    },
~~~h_C + h_B + h_A =\eta =\pm 1.
\ee
The formula describes all known splitting probabilities found earlier 
in QFT (\ref{setofquarkgluon}) and generalised Yang-Mills theory 
(\ref{setoftensorgluonpolariz1}), (\ref{setoftensorgluonpolariz2}). 
It describes splitting probabilities for integer and half-integer spin particles. 
This is a surprising and encouraging result because such a 
high symmetry was not 
explicitly implemented into the initial formulation. 
We have demonstrated that the splitting probabilities 
 (\ref{kernelsym1}) fulfil the generalised 
 Kounnas-Ross supersymmetry $N=1$ SUSY relations (\ref{tworelation})
hinting to the fact that the underlying theory can be formulated in an explicit supersymmetric  manner \cite{Antoniadis:2011re}. 

One of us, G.S., would like to thank the Institute of Theoretical Physics of the 
Leipzig University for hospitality, where part of this work was completed. 
This work was supported in part by the Alexander von Humboldt Foundation.   
G.S. was also supported in part by the Marie Sk\'lodowska-Curie Grant Agreement No 644121.

.

\end{document}